\documentstyle[preprint,aps,prl,eqsecnum,epsfig]{revtex}
\newcommand{\beq}{\begin{equation}}
\newcommand{\eeq}{\end{equation}}
\newcommand{\bea}{\begin{eqnarray}}
\newcommand{\eea}{\end{eqnarray}}

\newcommand\refpar[1]{(\ref{#1})}

\begin{document}
\draft
\title{Scattering of dislocated wavefronts
by vertical vorticity and the
Aharonov-Bohm effect I: Shallow water}
\author{{\sc Christophe Coste}$^a$, {\sc Makoto Umeki}$^{b}$, and
{\sc Fernando Lund}$^c$ \\[1em]
$^a$ Laboratoire de Physique, ENS Lyon\\
46, All\'ee d'Italie 69364 Lyon Cedex 07, France\\
$^{b}$Department of Physics, University of Tokyo, \\
 7-3-1 Hongo, Bunkyo-ku, Tokyo, 113 Japan \\
  $^{c}$Departamento de F\'\i sica, Facultad de
Ciencias F\'\i sicas y Matem\'aticas \\
 Universidad de Chile, Casilla 487-3,
Santiago, Chile}

\maketitle

\begin{abstract}
When a surface wave interacts with a vertical vortex in shallow water
the latter induces a dislocation in the incident wavefronts that is
analogous to what happens in the Aharonov-Bohm effect for the scattering
of electrons by a confined magnetic field. In addition to
this global similarity between these two physical systems there
is scattering. This paper reports a detailed calculation of this
scattering, which is quantitatively different from the electronic case
in that a surface wave penetrates the inside of a vortex while electrons
do not penetrate a solenoid. This difference, together with an
additional difference in the equations that govern both physical systems
lead to a quite different scattering in the case of surface waves, whose
main characteristic is a strong asymmetry in the scattering cross
section. The assumptions and approximations under which these effects
happen are carefully considered, and their applicability to the case of
scattering of acoustic waves by vorticity is noted.

\end{abstract}



\section{Introduction}
\label{sec:introduction}

In a remarkable paper, Berry et.  al.\cite{berry} clarified the way in
which a curl-free magnetic vector potential modifies the wavefront
structure of an electronic wavefunction that obeys the non relativistic
Schr\"odinger equation.  They concluded that for electrons travelling
outside an infinitely long cylinder enclosing a magnetic field, the
wavefronts outside the cylinder would be dislocated by an amount
proportional to the amount of magnetic flux within the cylinder.
Reasoning by analogy, they also concluded that such dislocated wavefronts
should occur for surface water waves when they encounter a vortex.  A
simple experiment conclusively demonstrated this effect\cite{berry}.

In the case of the electronic wavefunction interacting with a confined
magnetic field (and its unconfined vector potential) Berry et. al.
\cite{berry} also computed the complete solution to the Schr\"odinger
equation that, in addition to the dislocated wave, includes a scattered
wave. Trying to do this in the case of the water waves is however
more difficult because the analogy between de Broglie waves and water
waves breaks down when pushing it
into a quantitative statement. There are two essential differences:
The first is that for an
electron the appropriate boundary condition is that the wave function
vanishes at the surface of the cylinder; in the case of water
waves, the waves of course penetrate inside the vortex and it becomes
necessary
to solve the appropriate equations not only outside the vortex but
also inside, and match them with continuity conditions. The second is
that the wave equations that govern both phenomena, although similar,
differ in quantitative details. This paper addresses both these issues.

The scattering of surface waves by vertical vorticity in shallow water was
discussed by Cerda and Lund\cite{cerdalund} and by Umeki and
Lund\cite{umlu} who discovered that a vortex may support spiral wave
solutions.  Fabrikant and Raevsky\cite{fabray} have studied the case of a
fluid of arbitrary depth in a Born approximation.  The interaction between
surface waves and
vertical vorticity is in many respects similar to that of acoustic waves
and vorticity, a topic that has been much studied over the years and that
recently has been the subject of particular interest due to the
possibility of using acoustic waves as a nonintrusive probe of vortical
flows, both laminar and turbulent\cite{acoust+vort}, much in the same way
that X rays and neutrons are used to probe condensed matter structures,
both ordered and disordered.  Most treatments, however, rely on a Born
approximation\cite{fetter} whose validity breaks down when a surface
wave interacts with a vertical vortex with nonvanishing circulation,
leading to a long range velocity field that decays like $1/r$ where $r$
is the distance to the vortex core. This is precisely the case that is
studied in the present paper.

This paper is organized as follows: Section 2 derives the equations that
describe the scattering of a surface wave by vertical vorticity in shallow
water. We pay particular attention to the assumptions needed to
derive those equations. Section 3 has a reminder on the Aharonov-Bohm
effect as relevant for the present discussion. Section 4 presents the
computation of the solution to the equations derived in Section 2. Section
5 presents several illustrative examples and Section 6 has
concluding remarks. Technical details are contained in two Appendices. 
A subsequent paper\cite{deep} studies the first order corrections to the
shallow water approximation.

\section{Shallow water waves in interaction with a vertical vortex}
\label{sec:interaction}

We consider the  problem of the interaction of shallow water surface waves
in an  inviscid incompressible fluid of uniform depth $h$ with a
stationary vertical vortex. The coordinates are
 $(x,y)=\mbox{\boldmath $x$}$ in the horizontal direction and $z$ in the
vertical direction. The velocity and the surface displacement are denoted
by
$\mbox{\boldmath $v$}(\mbox{\boldmath $x$},z,t) = (\mbox{\boldmath
$v$}_\perp
(\mbox{\boldmath $x$},z,t),
w(\mbox{\boldmath $x$},z,t)) $ and $\eta(\mbox{\boldmath $x$},t)$,
respectively.

The equation of motion is\cite{book}
\begin{equation}
\partial_t \mbox{\boldmath $v$} + (\mbox{\boldmath $v$} \cdot \nabla)
\mbox{\boldmath $v$} = - \rho^{-1} \nabla p - \mbox{\boldmath $g$},
\label{s1}
\end{equation}
where $\rho$ is the (uniform) density of the fluid, $p$ is the pressure,
$\mbox{\boldmath $g$}$ is the gravitational acceleration, $\partial_t$ =
$\partial/\partial t$ and $\nabla$ the three-dimensional gradient.
The boundary conditions on $w$ are $w=0$ at the fluid's bottom ($z=0$), and
\begin{equation}
w= \partial_t \eta + \mbox{\boldmath $v$}_\perp  \cdot \nabla_\perp  \eta
\qquad (z = h + \eta(\mbox{\boldmath $x$},t) ) ,
\label{s2}
\end{equation}
where $\nabla_\perp $ is a horizontal gradient, at the surface. We
consider a free surface and neglect surface tension, which is consistent with
the shallow water approximation.

In shallow water the length scale for spatial variations is much bigger
than the fluid depth $h$. Consequently, the continuity equation
$\nabla \cdot \mbox{\boldmath $v$} =0$ together with the boundary
condition at the bottom imply
\begin{equation}
\label{eq:neweq}
\left. w (\mbox{\boldmath $x$}, z t) \right|_{z=h+\eta} =
-\nabla_\perp \cdot \left. \mbox{\boldmath $v$}_\perp \right|_{z=0}
(h+\eta)
\end{equation}
to leading order in $h/L$, where $L$ is the length scale for space
variations. Inserting (\ref{eq:neweq}) into the kinematic boundary
condition (\ref{s2})
and using that, to leading order in the shallow water approximation
\[
\left. \mbox{\boldmath $v$}_\perp \right|_{z=0}  =
\left. \mbox{\boldmath $v$}_\perp \right|_{z=h+\eta}
\]
leads to
\begin{equation}
\partial_t \eta + h \nabla_\perp \cdot \mbox{\boldmath $v$}_\perp  +
\nabla_\perp  \cdot (\eta \mbox{\boldmath $v$}_\perp ) =0 ,
\label{s9}
\end{equation}
assuming surface deformations small compared to depth ($\eta \ll h$).

Neglecting vertical accelerations with respect to $\mbox{\boldmath
$g$}$,
the $z$-component of (\ref{s1}) and continuity of the pressure at the free
surface yield
\begin{equation}
p(\mbox{\boldmath $x$},z,t) = \rho g (h + \eta(\mbox{\boldmath $x$},t) -z)
 + p_a,
\label{s7}
\end{equation}
where $p_a$ is the atmospheric pressure. Substitution into the
 $\mbox{\boldmath $x$}$-component of (\ref{s1}) gives, again in the
 shallow water approximation
\begin{equation}
\partial_t \mbox{\boldmath $v$}_\perp + (\mbox{\boldmath $v$}_\perp  \cdot
\nabla_\perp ) \mbox{\boldmath $v$}_\perp  = -g\nabla_\perp  \eta.
\label{s8}
\end{equation}

We will consider surface waves with particle velocity $\mbox{\boldmath
 $u$}
(\mbox{\boldmath $x$},t) $  and surface deformation
$\eta_1 (\mbox{\boldmath $x$},t) $ as small perturbations on a
background flow consisting of a steady vertical vortex $\mbox{\boldmath
$U$}(\mbox{\boldmath $x$})$, with corresponding
 surface deformation $\eta_0(\mbox{\boldmath $x$})$; $ u \ll U$, where $U$
denotes a typical value of $\mbox{\boldmath $U$}(\mbox{\boldmath $x$})
$, and
$\eta_1 \ll
\eta_0$. Substituting
$ \mbox{\boldmath $v$}_\perp  = \mbox{\boldmath $U$}(\mbox{\boldmath $
x$} ) +
\mbox{\boldmath $u$} (\mbox{\boldmath $x$},t) $ and
$ \eta = \eta_0(\mbox{\boldmath $x$}) + \eta_1 (\mbox{\boldmath $x$},t
) $ into (\ref{s8}) and (\ref{s9}) leads, to leading order in the small
perturbations $\eta_1$ and $\mbox{\boldmath $u$}$, to
\begin{equation}
-g\nabla_\perp  \eta_0  =  (\mbox{\boldmath $U$} \cdot \nabla_\perp )
\mbox{\boldmath $U$} = \frac{1}{2} \nabla_\perp  \mbox{\boldmath $U$}^
2 -
\mbox{\boldmath $U$} \times {\rm rot} \mbox{\boldmath $U$},
\label{s10}
\end{equation}
and
\begin{equation}
\mbox{\boldmath $U$} \cdot \nabla_\perp  \eta_0 =0 .
\label{s12}
\end{equation}
These equations allow the computation of $\eta_0$ in terms of $\mbox{\boldmath $U$}$
for the background flow. The first order equations are
\begin{equation}
\partial_t \mbox{\boldmath $u$} + (\mbox{\boldmath $U$} \cdot \nabla_\perp
)\mbox{\boldmath $u$} = - (\mbox{\boldmath $u$} \cdot \nabla_\perp )
\mbox{\boldmath $U$} - g \nabla_\perp  \eta_1,
\label{s14}
\end{equation}
\begin{eqnarray}
\partial_t \eta_1 + (\mbox{\boldmath $U$} \cdot \nabla_\perp  )\eta_1
 & = & -h\nabla_\perp  \cdot \mbox{\boldmath $u$}  \\ \nonumber
 & & - [ \eta_0\nabla_\perp  \cdot \mbox{\boldmath $u$} + (\mbox{\boldmath
$u$}
\cdot
\nabla_\perp ) \eta_0  ].
\label{s15}
\end{eqnarray}
Taking the horizontal divergence of (\ref{s14}), we obtain
\begin{equation}
\partial_t \nabla_\perp  \cdot \mbox{\boldmath $u$} + g\triangle_\perp
\eta_1
 = -\nabla_\perp  \cdot [(\mbox{\boldmath $U$} \cdot \nabla_\perp )
\mbox{\boldmath $u$} + (\mbox{\boldmath $u$} \cdot \nabla_\perp  )
\mbox{\boldmath $U$}],
\label{s16}
\end{equation}
where $\triangle_\perp = \nabla_\perp ^2$,
and rearranging the right hand side of (\ref{s16}) gives ($i,j = 1,2$)
 \begin{equation}
D_t \nabla_\perp  \cdot \mbox{\boldmath $u$} + g\triangle_\perp\eta_1
              = - 2(\partial_i U_j)(\partial_j u_i),
\label{s17} \end{equation}
where  $D_t \equiv \partial_t + \mbox{\boldmath $U$} \cdot \nabla_\perp $.
Taking the difference between $D_t$ of Eqn. (\ref{s15}) and $h$ times Eqn.
(\ref{s17}) leads to
\begin{equation}
D_t^2 \eta_1 - c^2 \triangle_\perp \eta_1 =
  - D_t [ \eta_0\nabla_\perp  \cdot \mbox{\boldmath $u$} +  (\mbox{
\boldmath
$u$}
\cdot \nabla_\perp ) \eta_0  ]
+ 2h(\partial_i U_j)(\partial_j u_i),
\label{eqpb}
\end{equation}
where $c= \sqrt{gh}$ is the phase velocity of shallow water waves.

We consider the case $ U \ll c$ .
In analogy with gas dynamics, we call $M=U/c$ the Mach number.
We denote a typical length scale of the vortex by $a$,
and the wavelength and frequency of shallow water waves by
$\lambda$ and $\nu$ respectively.
We will assume that wavelengths are small compared to vortex
size\cite{colonius}: $ k a \equiv
\beta \gg 1$ , ($k=2\pi/ \lambda$).

Under these assumptions, the right hand side of (\ref{eqpb}) will be
O($M$) or O($\beta^{-1}$) compared with the left hand side.
Neglecting these terms, the final equation to be solved is
\begin{equation}
D_t^2 \eta_1 - c^2 \triangle_\perp \eta_1 = 0.
\label{s19}
\end{equation}
Note that one might be tempted to neglect $(\mbox{\boldmath $U$} \cdot
\nabla_\perp ) \eta_1$ with respect to $\partial_t \eta_1$ on the
grounds that $U \ll c$. However, it is possible to have
$(\mbox{\boldmath $U$} \cdot
\nabla_\perp ) \eta_1 \sim \partial_t \eta_1$ without violating
this small Mach number assumption by considering (as we do in Section 4
below) a background velocity $ U \sim \omega a$ with length scale $a$
and frequency scale $\omega$, together with $\nu \gg \omega$ and $ka \gg
1$.

The equation \refpar{s19} is readily obtained, under the same assumptions, in
the diffusion of acoustic waves by a vortex \cite{umlu}. The physics is the same
since acoustic waves and shallow water waves are both {\it nondispersive}, and
the results of Sections\ \ref{sec:scattering} and
\ref{sec:examples} are valid for both types of waves. They depend only on two
parameters, the dimensionless wave number
$\beta$ and the Mach number $M$, and when those parameters are the same the
results that hold for the surface elevation $\eta_1$ may be transposed,
quantitatively and with no change, for the scattered acoustic pressure.

\section{Analogy with the Aharonov-Bohm effect}
\label{sec:analogy}

The wave equation (\ref{s19}) posesses a close analogy with the quantum
mechanical wave equation describing the Aharonov-Bohm effect, in which a
magnetic vector potential influences the dynamics of a charged particle
in a region where the magnetic field vanishes.  This cannot happen in
classical electrodynamics\cite{ced}.  In its simplest form, this effect
occurs when a beam of particles with charge $q$ and mass $m$ is incident
normally on a long thin cylinder containing a magnetic field
$\mbox{\boldmath $B$}(\mbox{ \boldmath $x$})$ parallel to its axis.  The
Schr\"odinger equation in the presence of a magnetic vector potential
$\mbox{\boldmath $A$}$ is
\begin{equation}
\frac{1}{2m}(-i\hbar\nabla - q \mbox{\boldmath $A$}(\mbox{\boldmath $x
$}) )^2
\psi(\mbox{\boldmath $x$})
= \frac{\hbar^2 k^2}{2m} \psi(\mbox{\boldmath $x$}),
\label{a1}
\end{equation}
where $\hbar$ is Planck's constant.
Outside the cylinder, $\mbox{\boldmath $A$}(\mbox{\boldmath $x$})=
(\Phi /2\pi r)
\hat{\theta}$, with $\Phi$ the magnetic flux contained within the
cylinder,
and $\hat{\theta}$ an azimuthal unit vector. Of course,
$\mbox{\boldmath $B$} = 0 $ outside the cylinder.

Both Equations (\ref{s19}) and (\ref{a1}) allow for a solution of the form
$$
\exp [-i(\mbox{\boldmath $k$} \cdot \mbox{\boldmath $x$} + \alpha \theta
)],
$$
where $\alpha = \nu \Gamma / (2 \pi c^2) = k \Gamma / 2 \pi
c$ in the fluid mechanics case and $\alpha = -q\Phi / 2\pi \hbar $ in the
quantum mechanics case.  This is an exact statement in the latter case,
while in the water wave case it is approximate, because (\ref{s19}) is
valid only when $M \ll 1$
and $\beta \gg 1$.  Except for integer values of $\alpha$, this is a
multivalued solution.  Berry et.  al.\cite{berry} showed how fixing this
multivaluedness leads to a solution that is a superposition of dislocated
wavefronts and scattered waves.  This was achieved by solving the
Schr\"odinger equation (\ref{a1}) with impenetrable boundary conditions:
$ \psi = 0 $ at the surface of the cylinder.  The appropriate boundary
conditions in the fluids case are continuity of velocity and of surface
elevation.  We now turn our attention to solving Eqn.  (\ref{s19}) under
these conditions.  One important physical difference between the classical
and
quantum mechanical cases is that in the latter the phase of the waves
cannot be measured, while in the classical case it can.  Table 1 compares
these two cases.

\section{Scattering of dislocated waves by a vortex}
\label{sec:scattering}

As an example, we consider a scattering problem by a circular uniform
vortex with vorticity $\omega$ and radius $a$ surrounded by
an irrotational flow.
Using polar coordinates $(r, \theta)$, the background flow is given
by\cite{foot1}
\begin{equation}
\mbox{\boldmath $U$} =
\left\{
\begin{array}{lcl}
\frac{1}{2} \omega r \hat{\theta}  &  & \mbox{if  $r\le a$} \\
 & & \\
\frac{\Gamma}{2\pi r} \hat{\theta} &  & \mbox{if  $r > a$}
\end{array}
\right.
\label{v1}
\end{equation}
where $\Gamma = \pi \omega a^2$ is the circulation.
Eqn. (\ref{s19}) will be solved separately for $r<a$ and $r>a$, and the
results matched with a continuity condition.

Inside the vortex we have, from (\ref{s19}),
\begin{equation}
[(\partial_t + (\omega/2) \partial_\theta)^2-c^2 (\partial_r^2 + (1/r)
\partial_r + (1/r^2) \partial_\theta^2)] \eta_1 =0.
\label{v2}
\end{equation}
We look for solutions that evolve harmonically (with a single global
frequency $\nu$) in time, and Fourier decompose them in the polar angle
$\theta$:
\begin{equation}
\displaystyle
\eta_1 = {\rm Re} [\sum_n \tilde{\eta}_{1n} {\rm e}^{i (n \theta - \nu
 t )} ] ,
\label{form}
\end{equation}
where Re stands for the real part. Introducing this expression into
(\ref{v1})
we obtain
\begin{equation}
\left(\frac{d^2}{d r^2} + \frac{1}{r} \frac{d}{d r} - \frac{n^2}{r^2}
+ k_n^2
\right)
\tilde{\eta}_{1n} =0, \qquad k_n = \frac{| \nu - n\omega/2 |}{c}.
\label{v3}
\end{equation}
Equation (\ref{v3}) has both Bessel and Neumann functions as solutions if
$k_n\ne 0$. Regularity at the origin will exclude the latter.
If $2\nu/\omega$ is an integer, $k_n (=k|1-n/n_d|)$ vanishes for $n=n_d
\equiv 2\nu/\omega$. In this case, (\ref{v3}) can be solved by assuming
$\tilde{\eta}_{1n} \propto r^p$. Substituting this into (\ref{v3}), we
have $p= \pm n$ and negative values of $p$ are excluded, again because of
regularity at the origin.  Thus we have
\begin{eqnarray}
\eta_1 (r, \theta, t) & = & {\rm Re} \left[ \sum_{n \ne n_d }
a_n \frac{J_{|n|}(k_n r)}{J_{|n|}(k_n a)} {\rm e}^{i(n\theta -\nu t)}
+ \right.
\\ \nonumber
 & & \left. C(n_d) a_{n_d} \left( \frac{r}{a} \right)^{n_d} {\rm e}^{i
(n_d \theta -\nu t)} \right],
\label{v4}
\end{eqnarray}
where the $a_n$ are as yet undetermined coefficients and $C(n_d) =
1$ when $2\nu /\omega$ is an integer and vanishes otherwise.

Outside the vortex, $r > a$, the assumption $U^2/c^2 \ll 1$ reduces
(\ref{s19}) to
\begin{equation}
\left[ \partial_t^2 + \frac{\Gamma}{\pi r^2} \partial_\theta \partial_
t -c^2
(\partial_r^2 + (1/r) \partial_r + (1/r^2) \partial_\theta^2) \right]
\eta_1 =0.
\label{v5}
\end{equation}
Inserting the form (\ref{form}) of $\eta_1$ into this equation gives
\begin{equation}
\left(\frac{d^2}{d r^2} + \frac{1}{r} \frac{d}{d r} -
\frac{n^2+2n\alpha}{r^2} +
k^2 \right)
\tilde{\eta}_{1n} = 0,  \qquad k = \frac{\nu}{c}.
\label{v6}
\end{equation}
with $\alpha = \nu \Gamma /2\pi c^2$. We wish this parameter to be of
order 1. Following Berry et al. \cite{berry} we write the surface
elevation outside the vortex in the form
\begin{equation}
\eta_1 = {\rm Re} (\eta_{AB}+\eta_R),
\label{v8} \end{equation}
where
\begin{equation}
\eta_{AB}= \sum_n b_n \frac{J_m(kr)}{J_m(\beta)} {\rm e}^{i(n\theta - \nu
 t)} ,
\qquad m \equiv  \sqrt{n^2 +2n\alpha },
\label{v9} \end{equation}
 with $\beta \equiv ka$, and
\begin{equation}
\eta_{R}= \sum_n  c_n \frac{H^1_m(kr)}{H^1_m(\beta)} {\rm e}^{i(n\theta -
 \nu t)}.
\label{v10} \end{equation}
The coefficients $a_n$, $b_n$ and $c_n$ are defined so that they
denote the amplitude of the wave components at the vortex boundary $r=a$.
In order to obtain these coefficients, the continuity of $\eta$ and
$\nabla_\perp  \eta$ at $r=a$ is required. This gives two relations:
\begin{equation}
a_n = b_n +c_n,
\label{v11} \end{equation}
\begin{equation}
a_n k_n \frac{J'_{|n|}(k_n a)}{J_{|n|}(k_n a)} =
k \left( b_n \frac{J'_m (\beta)}{J_m (\beta)}+c_n \frac{H^{1'}_m
(\beta)}{H^{1
}_m (\beta)}
\right).
\label{v12} \end{equation}
If $n= n_d$, the corresponding relations are
\begin{equation}
a_n = b_n +c_n ,
\label{v13} \end{equation}
\begin{equation}
a_n \frac{n}{a} = k \left( b_n \frac{J'_m (\beta)}{J_m (\beta)}
                    +c_n \frac{H^{1'}_m (\beta)}{H^{1}_m (\beta)} \right).
\label{v14}
\end{equation}

The third condition comes from the boundary condition of $\eta$ at
infinity.  We require that the asymptotics of $\eta_{AB}$ coincides with
the dislocated wave incident from the right plus outgoing waves only. This
leads to (see Appendix A)
\begin{equation}
\frac{b_n}{J_m(\beta)} = (-i)^m
\label{v15} \end{equation}

Using the notation
\begin{equation}
\gamma_n \equiv {k_n \over k} = \left|1 - {n\alpha \over \beta^2}\right|,
\label{v15bis}
\end{equation}
and the relationship $z Z'_{\nu}(z) =
zZ_{\nu -1}(z) - \nu Z_{\nu}(z)$, where $Z_{\nu}(z)$ is any one of the Bessel
functions, the following expressions   for  $a_n$ and
$c_n$ are obtained, when
$\gamma_n
\ne 0$ :
\begin{equation}
a_n = \frac{ (-i)^m J_m(\beta ) }{ \Delta_n} \left[-\frac{H_{m-1}^{1}(\beta
)}
{H_m^1(\beta )}+
        \frac{J_{m-1}(\beta) }{J_m(\beta)} \right] ,
 \label{v21} \end{equation}
\begin{equation}
c_n = \frac{(-i)^m J_m(\beta ) }{ \Delta_n} \left[-\frac{\gamma_n J_{
|n|-1}(\beta
\gamma_n)}
{J_{|n|}(\beta \gamma_n)} +\frac{J_{m-1}(\beta) }{J_m(\beta)} - {1\over
\beta}(m
- |n|)
\right] ,
\label{v22} \end{equation}
where
\begin{equation}
\Delta_n = -\frac{H^{1}_{m-1}( \beta )}{H^{1}_m( \beta )} +
\frac{\gamma_n J_{|n|-1}(\beta \gamma_n) }{J_{|n|}(\beta \gamma_n)}
+{1\over \beta}(m
- |n|),
                                         \ \ \ m= \sqrt{n^2 +2n\alpha}  .
\label{v23}
\end{equation}
If $\gamma_n=0$, i.e., $n=n_d$ , these formulae are replaced by
\begin{equation}
a_{n_d} = \frac{(-i)^{m_d} J_{m_d}(\beta )}{\Delta_{n_d}}
\left[-\frac{H_{{m_d}-1}^{1}(\beta) }{H_{m_d}^{1}
(\beta )}+
\frac{J_{{m_d}-1}(\beta)}{J_{m_d}(\beta)} \right] ,
\label{v24}
\end{equation}
\begin{equation}
c_{n_d} = \frac{(-i)^{m_d} J_{m_d}(\beta )}{\Delta_{n_d}} \left[-({n_d} +
m_d)
\beta^{-1} +
\frac{J_{{m_d}-1}(\beta)}{J_{m_d}(\beta)}
\right] ,
\label{v25}
\end{equation}
where
\begin{equation}
\Delta_{n_d} = -\frac{H^{1}_{{m_d}-1}( \beta )}{H^1_{m_d}(\beta)} + (m_d +
{n_d})
\beta^{-1} , \
\ \
{m_d}= \sqrt{{n_d}^2 +2{n_d}\alpha}  .
\label{v26} \end{equation}
These expressions are the main algebraic result of this paper.
\smallskip

The limit $ r \rightarrow \infty $ gives the surface elevation as
\begin{eqnarray}
\label{v16}
 \eta_{AB}  & \rightarrow & {\rm e}^{i(-kr \cos \theta + \alpha \theta
 - \nu t
)} \\ \nonumber
 & & - \frac{i{\rm e}^{i(kr-\nu t)}\sin \pi \alpha }{(2\pi i k r)^{1/2}
\cos(\theta /2 ) }
  (-1)^{[\alpha ]} {\rm e}^{ i([\alpha ]+1/2) \theta } \\ \nonumber
 & & + \frac{{\rm e}^{i(kr-\nu t)}}{(2\pi i k r)^{1/2}} G(\theta , -\pi /2)
,
\end{eqnarray}
where the function $G$ is defined in Appendix A, and $[\alpha]$ denotes the
integral
part of $\alpha$. The second term in the right hand side of the equation
diverges
for $\theta \to \pi$. This is because this asymptotics is valid
everywhere except in a narrow sector
centered around  the forward direction, $\theta = \pi$, of angular width
$O(1/\sqrt{kr})$,  where $\eta_{AB}$ cannot be separated into incident and
scattered waves \cite{singularity}, and it does not make sense to speak
of a forward scattering amplitude. This peculiarity
was already pointed out by Aharonov \& Bohm\cite{ced} in the case
of scattering by a {\it point} vortex.

Also
\begin{equation}
\label{v17}
\eta_R  \rightarrow \left(\frac{2}{\pi i k r}\right)^{1/2}
{\rm e}^{i(kr-\nu t)}
\sum_n {c_n \over H_m^1(\beta)}{\rm e}^{i(n\theta -\pi m/2)}.
\end{equation}
The sum of the last term of (\ref{v16})
and (\ref{v17}) is the correction to the Aharonov-Bohm scattering amplitude
that comes from the matching of the surface elevation and of its gradient
inside and
outside the vortex core.

Berry et al.
have calculated a correction for different boundary conditions. They
consider
the finite radius of the scattering solenoidal field, which is considered
as
impenetrable. In the quantum
mechanical context, the scattering is due to the magnetic field inside the
solenoid, and in an hydromechanical context it could be a solid body
rotating
in a perfect fluid. Their result reads
\cite{berry}
\begin{equation}
\label{v17bis}
\eta_R^{\rm Berry}  \rightarrow \left(\frac{2}{\pi i k r}\right)^{1/2}
{\rm e}^{i(kr-\nu t)}
\sum_n {J_{|n - \alpha|}(\beta) \over H_{|n - \alpha|}^1(\beta)}{\rm
e}^{i(n\theta -\pi |n - \alpha|)}.
\end{equation}

Since the usual scattering cross
section is not defined in the forward direction, it is interesting instead to 
compare the difference in
the far-field correction to the
Aharonov-Bohm wave function (obtained in the limit of zero cylinder
thickness) calculated by Berry et. al. on the basis of Schr\"odinger
equation, and our own calculations obtained on the basis of the fluids
equations. The general asymptotic form of
the
scattered wave
$\eta_S$ is
\begin{equation}
\eta_S \sim f(\theta) r^{-1/2} {\rm e}^{i(kr -\nu t)},
\label{v18}
\end{equation}
with a scattering amplitude $f(\theta)$ given by
\begin{equation}
f(\theta) = \frac{1}{\sqrt{2\pi i k}} \tilde{f} (\theta ).
\label{v19}
\end{equation}
In the following section, we will compare the correction to the Aharonov-Bohm
scattering amplitude for a vortex, that is
\begin{eqnarray}
\tilde{f}(\theta )  =   G(\theta , -\pi /2)  +
2 \sum_n \frac{c_n}{H_m^1(\beta)}  {\rm e}^{in \theta }(-i)^m,
             \label{v20}
\end{eqnarray}
with the correction for an impenetrable solenoidal field,
\begin{eqnarray}
\tilde{f}_{\rm Berry}(\theta )  =   2\sum_n {J_{|n - \alpha|}(\beta) \over
H_{|n - \alpha|}^1(\beta)}{\rm e}^{i(n\theta -\pi |n - \alpha|)}.
             \label{v20bis}
\end{eqnarray}

\section{Numerical Examples}
\label{sec:examples}

The solutions we have obtained are parametrized by two
dimensionless numbers: $\alpha  = \nu \Gamma /2\pi c^2$ and
$\beta = k a$. That is, for a given incident wave, they depend on vortex
radius and circulation as independent parameters. The Mach number is
related to $\alpha $ and $\beta$ through $\alpha= M \beta$.
Scaling radial distance with the vortex radius, $r' \equiv r/a$,
the analytical expression of the surface displacement is summarized as
follows:
$$\eta_1 = {\rm Re} \ \eta_c, \quad 0<r'\le 1 $$
\begin{eqnarray}
\eta_c & = & \sum_{n \ne n_d} a_n \frac{J_{|n|}(\gamma_n \beta
r')}{J_{|n|}(\gamma_n
\beta )}
 {\rm e}^{i(n\theta -\nu t)}  \\ \nonumber
 & & +C(n_d) a_{n_d} r'^{n_d}  {\rm e}^{i(n_d\theta -\nu
t)},
\label{v27} \end{eqnarray}
$$\eta_1 = {\rm Re} (\eta_{AB} + \eta_{R}), \ \ r'>1 $$
\begin{equation}
\eta_{AB} = \sum_n (-i)^m J_m(\beta r') {\rm e}^
{i(n\theta -\nu t)},
\label{v28} \end{equation}
\begin{equation}
\eta_R = \sum_n c_n \frac{H^1_m (\beta r')}{H^1_m (\beta)}
{\rm e}^{i(n\theta -\nu t)}.
\label{v29}
\end{equation}
where $m = \sqrt{n^2 +2n\alpha}$.

We have numerically computed the total surface displacement given by
(\ref{v27}-\ref{v29})  for several values of the parameters $\alpha$ and
$\beta$. In order to approximate the series in (\ref{v27}-\ref{v29})
by a finite sum, it is necessary to estimate their convergence. This is
done in
Appendix B, where it is shown that $\eta_c$ is an absolutely and uniformly
convergent series, and that $\eta_{AB}$ and $\eta_R$ are both absolutely
and
simply convergent series. As an illustration, absolute values of the
coefficients $a_n$ and $c_n$ are plotted in Fig. 1.

Since convergence of the series expansions for $\eta_{AB}$ and $\eta_R$ is
not
uniform, the number of terms to keep in the infinite series depends on the
value
of $r'$. In practice, we compute the patterns of the surface displacement
in the
region $|x'|, |y'| \le 5  [ (x',y')=(r'\cos \theta, r' \sin \theta )]$ by
the
finite  sum of (\ref{v27}) and (\ref{v29}) with $|n| \le 50$ for $\beta
=10$ and
$|n|
\le 30 $ for $\beta =5$, but we keep more terms, $|n|
\le 90 $ in \refpar{v28}. Fig. 2 shows the
resulting displacements for $\beta = 5$ and $\alpha = 0.5,\,1,\,1.5,\,2$,
and Fig. 3 for
$\beta = 10$ and the same values of $\alpha$. The dislocation of the
incident
wavefronts by an amount equal to $\alpha$ is clearly visible. The outward
travelling  scattered wave is also visible. Note the strong interference
patterns between scattered and incident wave.

Another illustration is given in
Fig. 4, where we substract to the total field the dislocated wave. The
scattered
wave appears as an outgoing cylindrical wave, with a
clearly visible dislocation in the forward
direction. This is the part of the wave that {\it does
not}
decrease as $1/\sqrt{kr}$, and that ensures single-valuedness of the total
field.
Note that the representation is for an half-integer value of $\alpha$, and
the scattering amplitude is exactly zero in the direction $\theta = \pi$
\cite{berry}; the comparison between the two figures clearly shows the
exact
compensation of the dislocation in this direction, because of destructive
interference, to yield a single valued total wave field.

Finally, Fig. 5 shows the absolute value of the correction to the Aharonov-Bohm
scattering amplitude, compared to the correction calculated by Berry et al.
For
$\alpha \geq 0.5$, the parameter $m$ is imaginary for small negative $n$. This
induces very different partial amplitudes for $\exp (-in\theta)$ and
$\exp (in\theta)$ when $n$ is small.
Our calculations thus predict a forward
scattering
with a strong asymmetry, which increases with $\alpha$ as shown
in Figs. 5 (c) and 5 (d). This asymmetry effect is
observed both in experiments on water wave scattering by a vortex
\cite{vivancomelo}
and in direct numerical simulations of sound scattering by a vortex
\cite{berthet}. As can be seen from the dashed curves in Fig. 5, this asymmetry
is absent in the calculation of Berry et al. For $\alpha \le 0.5$, the
parameter $m$ is real for all $n$ and the scattering in the forward
direction is much less asymmetric (Figs. 5 (a) and 5 (b)).

All calculations were performed
using  Mathematica\cite{mathematica}.

\section{Concluding Remarks}
\label{sec:Conclusion}

We have computed the surface displacement due to a surface wave
interacting with a vertical vortex in shallow water when the vortex core
performs solid body rotation, the wavelength is small compared
to the vortex core radius and the particle velocities associated with the
wave are small compared with the particle velocities associated with the
vortex. When the parameter $\alpha = \nu \Gamma /2 \pi c^2$ is of order
one or bigger, the wavefronts become dislocated. The scattered waves
interact strongly with the dislocated wavefronts and produce interference
patterns. The differential scattering cross section is strongly peaked
along a direction at an angle with respect to the incident direction. This
is in contrast with previous calculations of Berry et al. \cite{berry} in
the case of quantum mechanical scattering by an impenetrable 
cylinder of finite radius. In the sequel to this paper\cite{deep}, 
we will show that these properties roughly persist when
the depth of  the water increases. This is important because 
deep water waves are much more amenable to actual experiments.

\acknowledgements 

The work of F.L. is supported by Fondecyt
Grant 1960892 and a C\'atedra Presidencial en Ciencias. We gratefully
acknowledge a grant from ECOS-CONICYT.


\appendix
\section{Asymptotics}

In this Appendix, we study the asymptotic behavior of the function
\refpar{v9}. To this end,
we use the computations in the Appendix of Berry et. al.'s
paper\cite{berry}.  In order to avoid
confusion, we use the following notations: Our
definition of $m$ is called $m_{\rm new} \equiv \sqrt{n^2 +2n\alpha }$,
whereas the function used in \cite{berry} is called $m_{\rm old} \equiv
|n+\alpha
|$. Similarly, we note ${\tilde b}_n$ the constants in the series
representation of $\eta_{AB}$ in our work, whereas the constants for
$\eta_{AB}^{{\rm Berry}}$ are noted $b_n$.

Solutions to our Eqn. (\ref{v6}) are
Bessel functions of order
$m_{\rm new}$.  Moreover, the dislocated wave
\begin{equation}
\label{eq:dislocwf}
\exp[-i(\vec k \cdot \vec x + \alpha \theta +\nu t )]
\end{equation}
is a solution of Eqn.  (\ref{s19})
asymptotically, that is for $kr \gg \alpha$.  Consequently, it is
appropriate to take as a boundary condition at large distances
from the vortex that the solution should approach this dislocated wave.

Let us consider
\[
\eta_{AB}= \sum_n {\tilde b}_n \frac{J_{m_{\rm new}} (kr)}{J_{m_{\rm new}}
(\beta)}
{\rm e}^{i(n\theta - \nu t)}
\]
Coefficients ${\tilde b}_n$ should be determined from the boundary
condition that
$\eta_{AB}$ should tend asymptotically to (\ref{eq:dislocwf}) plus
purely outgoing cylindrical waves.
The representation
\begin{equation}
\label{eq:bessel}
J_m (z) = \frac{1}{2\pi}\int_{-\pi + i\infty}^{\pi + i\infty} {\rm e}^{i(mt - z
\sin t )} dt
\end{equation}
is still valid for $m = m_{\rm new}$, even for those $m$'s that
are purely imaginary (Ref.
\cite{gr} p. 954, formula 8.412.6). This happens when $\alpha$
is bigger than 0.5, and for those $n$'s satisfying
\[
-2\alpha < n < 0 .
\]

Next, we note that as $n$ grows, with $\alpha \sim O(1)$, the
difference between $m_{\rm old}$ and $m_{\rm new}$
decreases rapidly. Consequently, there will be an $N$, such that,
if $n>N$, the difference between the two $m$'s will be smaller than any
preasigned value. Let us write
\begin{eqnarray}
\eta_{AB}= \eta_{AB}^{\rm point} +
\sum_{|n|<N} \left( {\tilde b}_n
\frac{J_{m_{\rm new}} (kr)}{J_{m_{\rm new}} (\beta)} -
b_n \frac{J_{m_{\rm old}} (kr)}{J_{m_{\rm old}} (\beta)} \right)
 {\rm e}^{in\theta}  + \nonumber\\
+ \underbrace{\sum_{|n|>N} \left( {\tilde b}_n
\frac{J_{m_{\rm new}} (kr)}{J_{m_{\rm new}} (\beta)} -
b_n \frac{J_{m_{\rm old}} (kr)}{J_{m_{\rm old}} (\beta)} \right)
 {\rm e}^{in\theta}}_{\equiv R_N}.
\label{decomp}
\end{eqnarray}
The wave $\eta_{AB}^{\rm point}$ is the original result of Aharonov \& Bohm
\cite{ced}, and represents the scattering by a {\it point} vortex,
hence the notation. The decomposition \refpar{decomp} is interesting only
if the
last sum,
$R_N$, is small when $N$ is sufficiently large. We will see that it is
indeed the
case, and we temporarily drop it from the calculations.

We know that if
\[
\frac{b_n}{J_{m_{\rm old}}(\beta)} = (-i)^{m_{\rm old}}
\]
then $\eta_{AB}^{\rm point}$ gives a dislocated wavefront plus an
outgoing cylindrical wave. Next, if
\[
\frac{{\tilde b}_n}{J_{m_{\rm new}}(\beta)} = (-i)^{m_{\rm new}}
\]
 we may write,
\[
\eta_{AB} = \eta_{AB}^{\rm point} + \int_{-\pi + i\infty}^{\pi + i\infty} dt
{\rm e}^{-ikr \sin t } G(\theta ,t)
\]
where
\begin{equation}
\label{eq:g}
G(\theta ,t) \equiv \sum_{|n| < N} {\rm e}^{in\theta } \left( {\rm
e}^{im_{\rm new} (t- \pi
/2)} - {\rm e}^{im_{\rm old} (t- \pi /2)} \right)
\end{equation}
is an analytic function of $t$, because it is a finite sum of
analytic functions (exponentials). Also, it is dominated by the
contribution from low  $n$'s. For $kr  \rightarrow \infty$,  $\eta_{AB}$
can still be evaluated using  steepest descent.  Since $G$ does not have
any poles, the pole contribution to  $\eta_{AB}$ is the  same as that of
Berry et. al., namely Eqn. (A4) of \cite{berry}. This is good, since it is
just
the  dislocated
incident wave. Also, $G(t=\pi/2) = 0$ for all  $\theta$, including the
forward and backwards directions. This means that there are no  further
contributions from the $t= \pi /2$ saddle point. This is also good, since
the
outgoing character of the scattered wave is preserved. On the
other hand,
\[
G(\theta, -\pi /2) = \sum_{|n| < N} {\rm e}^{in\theta } \left( {\rm
e}^{-im_{\rm new}
\pi } - {\rm e}^{-im_{\rm old} \pi} \right) .
\]
This is different from zero for all $\theta$, including the
forward and  backwards directions and we have, outside a small angular
sector around the forward direction, the following asymptotic behaviour
at large distances:
\begin{equation}
\eta_{AB} (r \rightarrow \infty ) = \eta_{AB}^{\rm point} (r \rightarrow
\infty )
+ \frac{e^{ikr}}{\sqrt{2\pi i kr}} G(\theta, -\pi /2)
\label{saddle}
\end{equation}
This result differs from that obtained by Berry et. al.\cite{berry} on the
basis of Schr\"odinger's equation.

Let us turn back to the behavior of $R_N$ at large $N$. We consider the
behavior of
\begin{equation}
|R_N| < \sum_{|n|>N} |(-i)^{m_{\rm new} - m_{\rm old}}J_{m_{\rm new}} (z) -
J_{m_{\rm old}} (z)|
\label{RNdepart}
\end{equation}
where $z = kr$ is a fixed number. Using the asymptotic expressions of
Bessel
functions for large values of the index (Ref.
\cite{gr}, formula 8.452.1), we have
\begin{equation}
J_{|n + \alpha|}(z) \sim {e^{|n + \alpha|(\tanh \delta_1 - \delta_1)}\over
\sqrt{2\pi |n
+
\alpha| \tanh \delta_1}}, \qquad |n + \alpha| \equiv z \cosh \delta_1,
\label{BesselAsymptotique a}
\end{equation}
\begin{equation}
J_{\sqrt{n^2 + 2n\alpha}}(z) \sim {e^{\sqrt{n^2 + 2n\alpha}(\tanh \delta_2
-
\delta_2)}\over
\sqrt{2\pi \sqrt{n^2 + 2n\alpha} \tanh \delta_2}}, \qquad \sqrt{n^2 +
2n\alpha}
\equiv z \cosh \delta_2,
\label{BesselAsymptotique b}
\end{equation}
where $\sim$ means that we consider only the dominant behavior at large
$n$. An
important point is that these expressions suppose that $n>z$. The following
study
concerns simple convergence of the series $R_N$, for a fixed value of $z$,
not
uniform convergence valid for all $z$. We define
\[
\epsilon = {\alpha ^2 \over 2 z |n + \alpha|} = O(1/n),
\]
so that large $n$ behavior means small $\epsilon$.
It is easy to show that $ \delta_2 \sim \delta_1 - \epsilon/\sinh
\delta_1$, and that
$\sqrt{n^2 + 2n\alpha} \sim |n + \alpha| - \epsilon z$. We then deduce that
\[(-i)^{m_{\rm new} - m_{\rm old}} \sim 1 + i{\pi \over 2}\epsilon z,
\]
\[
\sqrt{n^2 + 2n\alpha}(\tanh \delta_2 -
\delta_2) \sim |n + \alpha|(\tanh \delta_1 - \delta_1) + \epsilon z
\delta_1,
\]
and that
\[
\sqrt{n^2 + 2n\alpha}\tanh \delta_2 \sim |n +
\alpha| \tanh \delta_1 - \epsilon z \cosh\delta_1/\sinh\delta_1,
\]
\[
\delta_1 = {\rm Argcosh}{|n + \alpha| \over z} = O(\log n),
\]
so that
\[
\epsilon z \delta_1 = O(\log n /n) \ll 1
\]
for large $n$. We have thus
\[
(-i)^{m_{\rm new} - m_{\rm old}}J_{\sqrt{n^2 + 2n\alpha}}(z) - J_{|n +
\alpha|}(z) \sim \epsilon z (\delta_1 + i\pi/2) J_{|n
+ \alpha|}(z) = O\left({\log n \over n}{n^{-n} \over \sqrt{n}}\right)
\]
Using only the very rough inequality $\log n / n \sqrt{n} <1$, we
can now conclude on the asymptotic behavior of $R_N$ at large $N$. Begining
with
\refpar{RNdepart}, we obtain
\begin{equation}
|R_N|< \sum_N^{\infty} N^{-n} < N^{-N}
\end{equation}
up to prefactors that we have dropped. The important point is that, indeed,
$R_N$
is a very small correction at large $N$, which validates the preceeding
analysis.
As a last remark, we insist on the fact that all calculations are done for
a {\it
fixed value} of $z$, and that $N$, at a prescribed accuracy, may depend on
$z$.

\section{Convergence}

In this appendix, we discuss the convergence of the numerical series
(\ref{v27}-\ref{v29}).

The simplest case is that of $\eta_{AB}$. In this case, the variable $r'$
may
extend toward infinity, and we fix its value in the calculations.
Therefore we
can conclude only on simple convergence of the series, not uniform
convergence. For
large
$n$,
$m
\sim n$. Using the formula
\refpar{BesselAsymptotique a}, and for a {\it fixed} value of $z' \equiv
\beta r'$, the angle $\delta = O(\log n)$ and we get
\[
J_m(z') \sim {1\over \sqrt{n}}\left({e\over n}\right)^n,
\]
so that most clearly the series \refpar{v27} is absolutely simply
convergent.

In the coefficients $a_n$ and $c_n$, some functions depends on $\gamma_n
\beta$,
and from \refpar{v15bis} we get $\gamma_n \beta \sim n M$ where $M \ll 1$ is the
Mach
number. Thus $\gamma_n \beta \ll n $, so that to get the asymptotic
behavior
at large $n$ of $J_n(\gamma_n \beta)$ we use the same formula
\refpar{BesselAsymptotique a} as before, but the angle $\delta$ is now a
constant
of order one. We then deduce the asymptotic behavior of
$\Delta_n$ from its expression
\refpar{v23}
\begin{equation}
\Delta_n = -\underbrace{\frac{H^{1}_{m-1}( \beta )}{H^{1}_m( \beta
)}}_{=O(1)} +
\underbrace{\frac{\gamma_n J_{|n|-1}(\beta \gamma_n) }{J_{|n|}(\beta
\gamma_n)}}_{=O(n)} +\underbrace{{1\over
\beta}(m - |n|)}_{=O(1/n)} = O(n).
\end{equation}
We have seen in the preceeding paragraph that the convergence of
$J_m(\beta)$ is
extremely fast, which ensures convergence of $a_n$. For $0\le
r' \le 1$, the term $J_{|n|}(\gamma_n \beta r')/J_{|n|}(\gamma_n \beta )$
takes the maximum value at $r'=1$ for sufficiently large values of $n$.
Then the absolute convergence of the sum \refpar{v27} is guaranteed by the
absolute convergence of the  coefficients $a_n$. Since the support of
$\eta_c$
is compact, this convergence is uniform.

Rather easily, we get that the asymptotic behavior of $c_n$ is that of
$J_m(\beta)$, which converges very rapidly. Let us introduce
\[
\beta r' \equiv m/\cosh \delta_1, \quad \beta  \equiv m/\cosh \delta_2.
\]
We have that $\delta_1 < \delta_2$, both being asymptotically of order $\log n$.
Using
one more time the formula \refpar{BesselAsymptotique a}, and its equivalent
for
Neumann functions (Ref. \cite{gr}, formula 8.452.2), we get
\[
\frac{H^{1}_{m}( \beta r')}{H^{1}_m( \beta )} \sim {{ E_1  - i  F_1}
\over { E_2  - i  F_2 } }\sim \exp[m(\delta_1 - \delta_2)]
\]
where ($i=1,2$)
\begin{eqnarray*}
E_i & \equiv & \exp(m\tanh \delta_i - m \delta_i) \over \sqrt{2\pi m\tanh
\delta_i } \\
F_i & \equiv & \exp(m \delta_i - m\tanh \delta_i )\over \sqrt{\pi m\tanh
\delta_i/2 }
\end{eqnarray*}
which converges exponentially fast because $\delta_1 - \delta_2 < 0$. We
deduce
that $\eta_R$ is an absolutely converging series. However, in this case,
$r'$ takes
values in an infinite interval so that the convergence is only simple.



\begin{figure}
\caption{Plot of the absolute value of the coefficients $a_n$ (a) and $c_
n$ (b) versus $n$ in a log-linear scale
for $(\alpha, \beta) =(0.5,10)$, denoted by dots, $(\alpha,
\beta) =(1.5,10)$, denoted by empty circles,
$(\alpha, \beta) =(0.5,5)$, denoted by filled circles
 and $(\alpha, \beta) =(1.5,5)$, denoted by empty squares. Note the
asymmetry with respect to $n \rightarrow -n$ }
\label{figure-1}
\end{figure}

\begin{figure}
\caption{Density plot of the surface elevation for the total wave patterns
for
$\beta = 5$, $\alpha = 0.5$ : (a), $\alpha = 1$ : (b), $\alpha = 1.5$ :
(c),
$\alpha = 2$ : (d). The greyscale is linear with surface  amplitude
(arbitrary
units). The dark circle  indicates the vortex
location. Vortex rotation is counterclockwise. The box size is 10
$\times$ 10 in units of the vortex radius $a$.
The incident wave comes from the right edge of the box. Note the
dislocated wave, and the asymmetric scattering that occurs practically
within a single quadrant.}
\label{figure-2}
\end{figure}

\begin{figure}
\caption{Same as figure 3, for $\beta = 10$, $\alpha = 0.5$ : (a), $\alpha
= 1$ :
(b),
$\alpha = 1.5$ :  (c), $\alpha = 2$ : (d).}
\label{figure-3}
\end{figure}

\begin{figure}
\caption{Density plot of the surface elevation for a dislocated incident
wave,
with parameters $\alpha = 1.5$, $\beta = 5$(resp. $\beta = 10$) : (a)[resp (c)],
and for the difference between
the total wave field with the same parameters, represented in Fig 2
(c) [resp. Fig 3 (c)] and
the dislocated wave : (b)[resp. (d)]. Figures (b) and (d) correspond to
the scattered wave generated by an incident dislocated wave. Such a
scattered wave is itself dislocated in the forward direction, thus
ensuring single valuedness. }
\label{figure-4}
\end{figure}

\begin{figure}
\caption{Polar plot of the absolute value
of the correction to the Aharonov-Bohm (i.e. point)
scattering amplitude, in the case of an impenetrable cylinder (dashed line)
and
in the case of a vortex (solid line), for $\beta = 10$ and $\alpha = 0.25$
: (a),
$\alpha = 0.5$ : (b), $\alpha = 1$ :  (c),
$\alpha = 1.5$ : (d). }
\label{figure-5}
\end{figure}

\clearpage
\newpage

\begin{table}
\caption{Aharonov-Bohm effect in quantum and classical mechanics compared
and contrasted.}
\label{tableone}
\begin{center}
\begin{tabular}{|rl|}
  Quantum mechanics    & Fluid mechanics             \\ \hline
  magnetic field & vorticity            \\ 
$\mbox{\boldmath $B$} = \nabla \times \mbox{\boldmath $A$}$ &
$\mbox{\boldmath $\omega$} = \nabla \times \mbox{\boldmath $U$}$ \\ \hline
vector potential $\mbox{\boldmath $A$}$ & velocity $\mbox{\boldmath $U$}$
 \\ \hline
magnetic flux $\Phi$ & velocity circulation $\Gamma$ \\ \hline
wave function $\psi$ & surface displacement $\eta$ \\ \hline
dislocation parameter & dislocation parameter  \\
$\alpha = -q\Phi/2 \pi \hbar $ & $ \alpha = k \Gamma /2 \pi c $ \\ \hline
dislocated wave  is      &  dislocated wave  is \\
an exact solution & approximate solution  \\ \hline
phase is  & phase is \\
not measurable & measurable \\
\end{tabular}
\end{center}
\end{table}



\clearpage
\newpage

\begin{figure}
\vskip -2truecm
\epsfysize=160mm
\centerline{\epsfbox{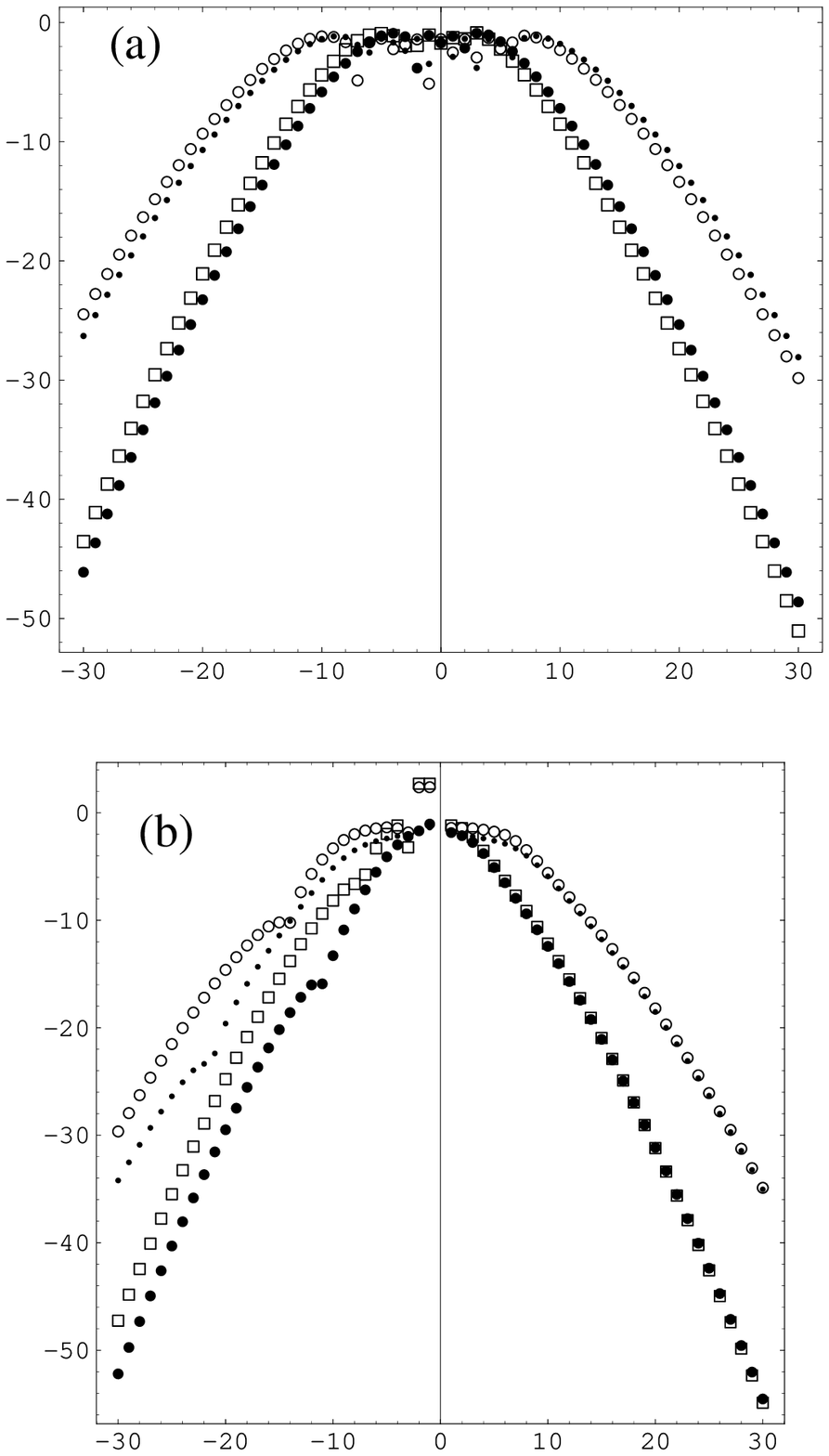}}
\vskip 3truecm
C. Coste {\it et al.}, Figure \ref{figure-1}
\end{figure}

\newpage

\begin{figure}
\epsfysize=120mm
\centerline{\epsfbox{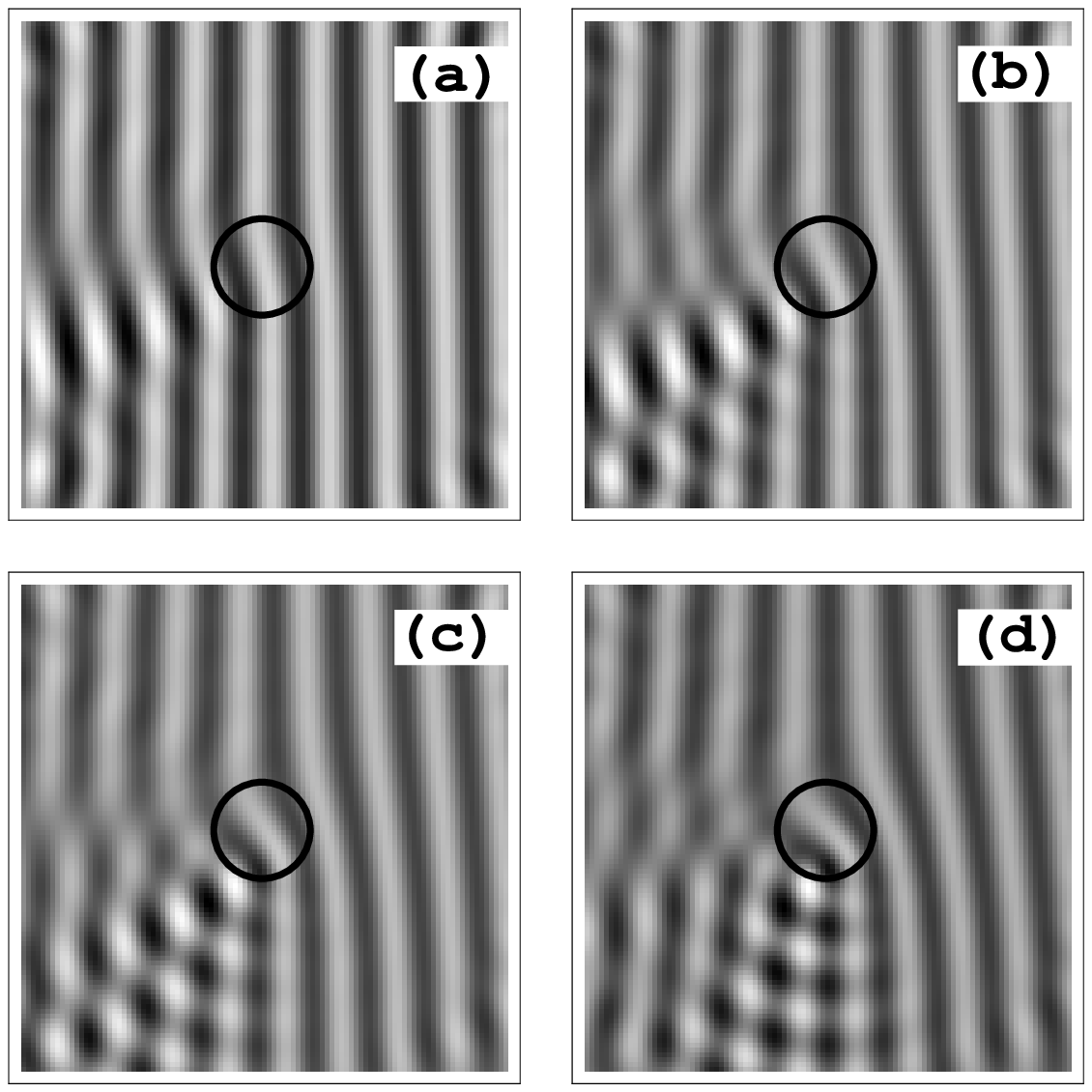}}
\vskip 3truecm
C. Coste {\it et al.}, Figure \ref{figure-2}
\end{figure}

\newpage

\begin{figure}
\epsfysize=120mm
\centerline{\epsfbox{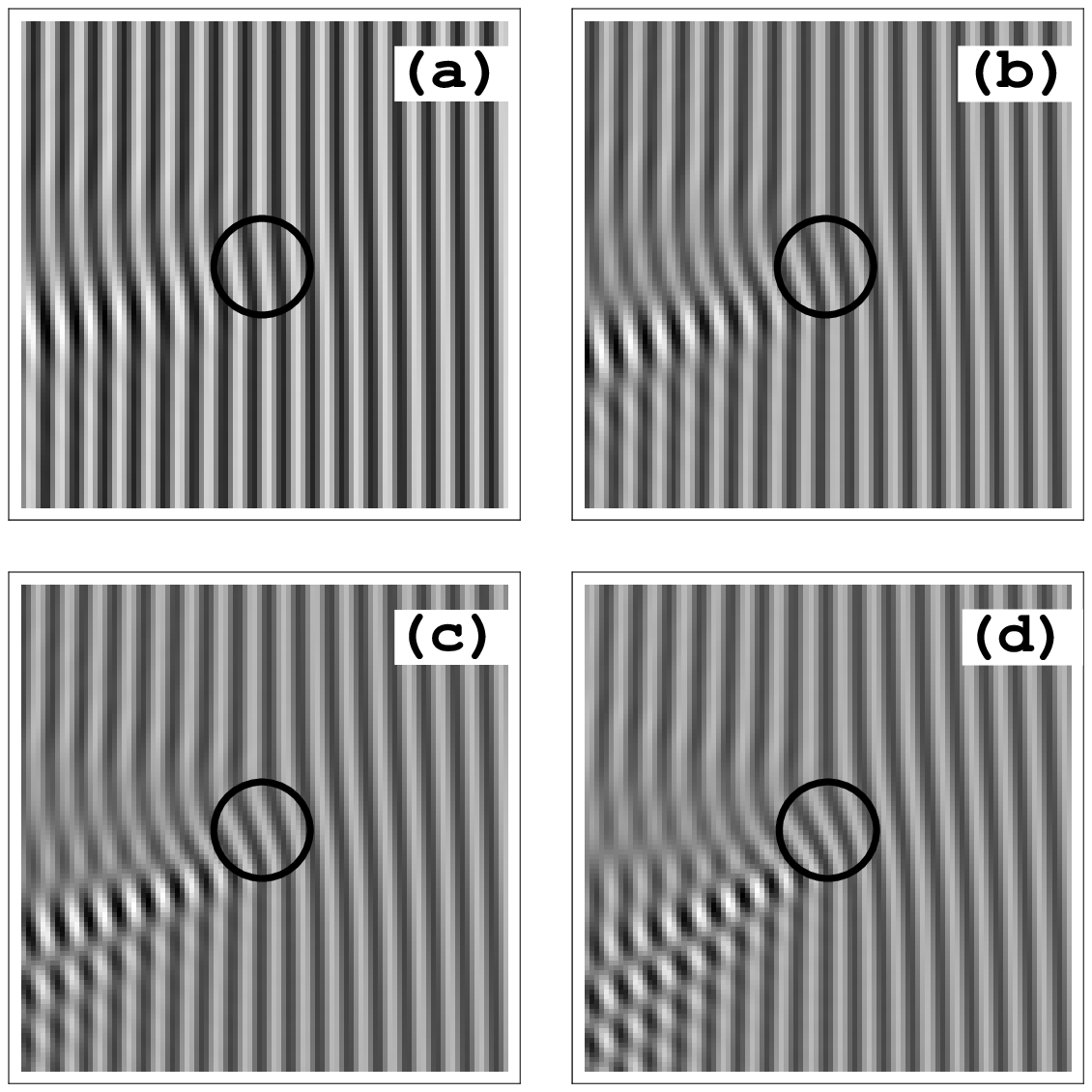}}
\vskip 3truecm
C. Coste {\it et al.}, Figure \ref{figure-3}
\end{figure}

\newpage

\begin{figure}
\epsfysize=120mm
\centerline{\epsfbox{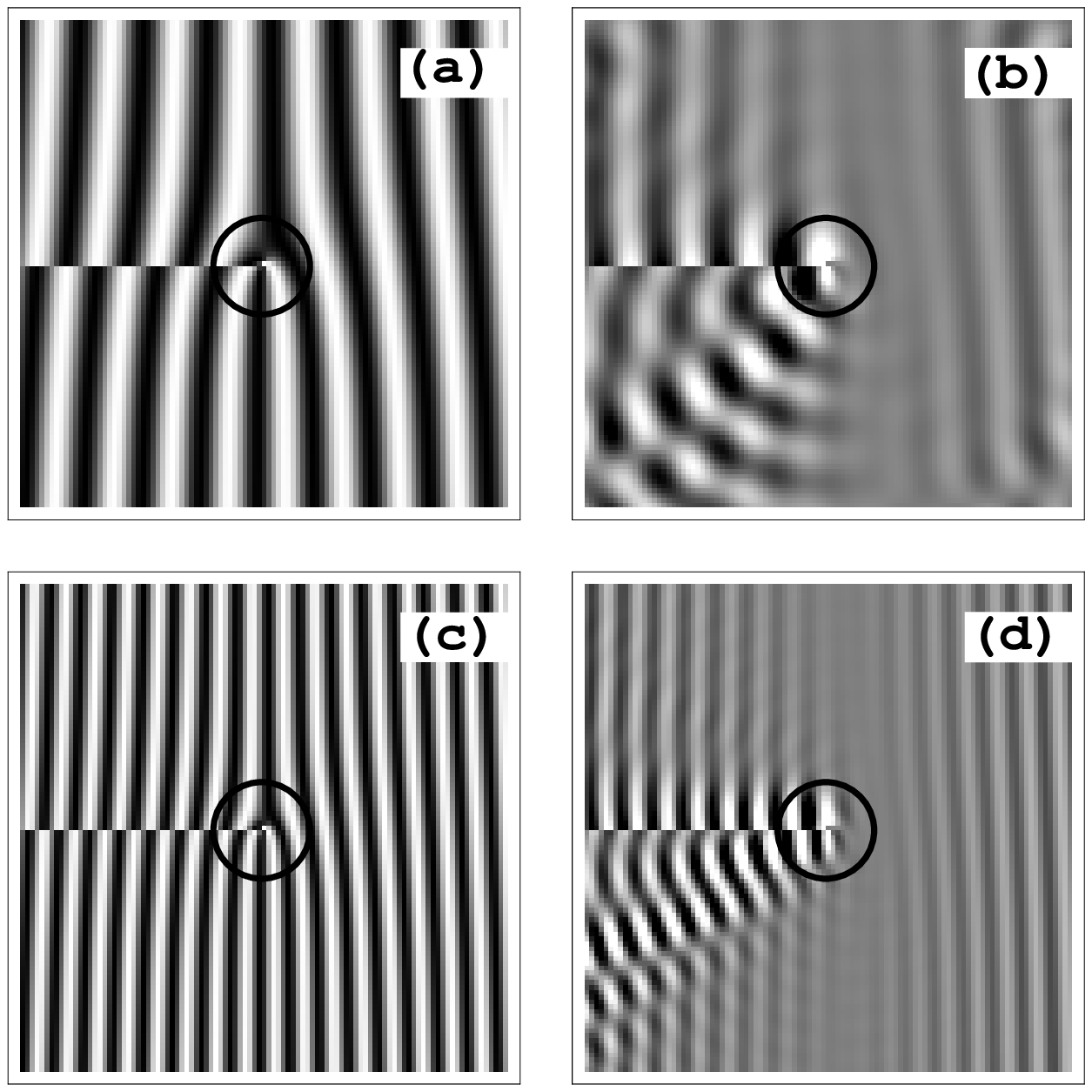}}
\vskip 3truecm
C. Coste {\it et al.}, Figure \ref{figure-4}
\end{figure}

\newpage

\begin{figure}
\epsfysize=120mm
\centerline{\epsfbox{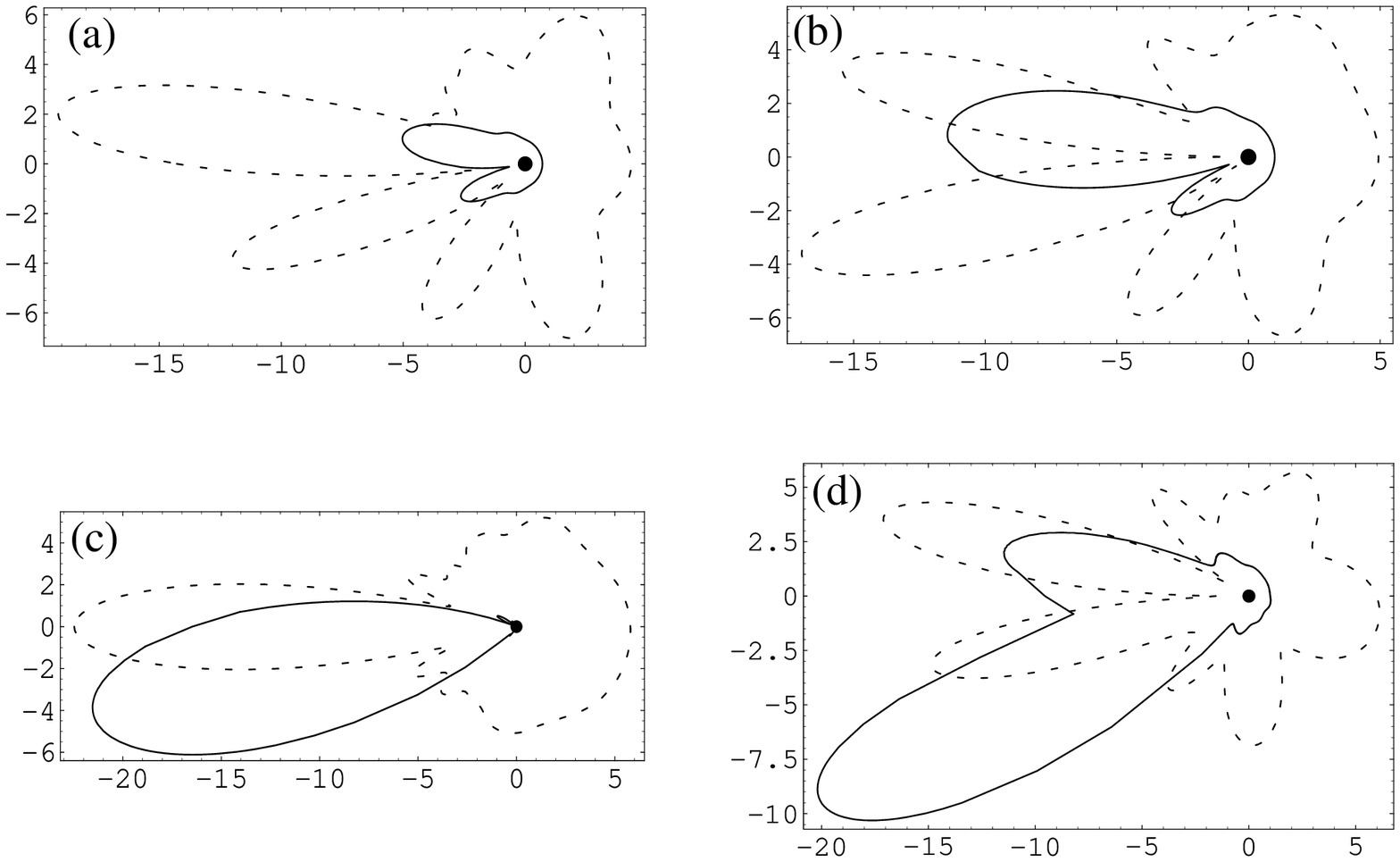}}
\vskip 3truecm
C. Coste {\it et al.}, Figure \ref{figure-5}
\end{figure}


\begin{references}
\bibitem{berry} M. V. Berry et. al., {\it Eur. J. Phys.} {\bf 1}, 154
(1980).
\bibitem{cerdalund}  E. Cerda and F. Lund, {\it Phys. Rev. Lett.} {\bf
70}, 3896  (1993).
\bibitem{umlu} M. Umeki and F. Lund, {\it Flu. Dyn. Res.} {\bf 21}, 201
(1997).
\bibitem{fabray} A. L. Fabrikant and M. A. Raevsky, {\it J. Fluid Mech.}
{\bf 262}, 141 (1994)
\bibitem{acoust+vort}
B. Dernoncourt, J.-F. Pinton and S. Fauve, {\it Physica D}, to appear;
M. Oljaca et. al. {\it Phys. Fluids A}, {\bf 10}, 886 (1998);
A. Petrossian and J.-F. Pinton, {\it J. de Physique II (France)} {\bf
7}, 1 (1997); J. F. Pinton et. al., {\it J. de Physique II (France)} {\bf
3}, 3 (1993)
C. Baudet, S. Ciliberto and J. F. Pinton, {\it Phys. Rev. Lett.} {\bf
67}, 193 (1991); H. Contreras and F. Lund, {\it Phys. Lett. A} {\bf 149},
127 (1990); F. Lund and C. Rojas, {\it Physica D} {\bf 37}, 508 (1989); M.
S. Howe, {\it J. Sound Vib.} {\bf 87}, 567 (1983) T. Kambe, {\it J. Japan
Soc. Fluid Mech.} {\bf 1}, 149 (1982) (in japanese); P. R. Gromov, A. B.
Ezerskii and A. L. Fabrikant, {\it Sov. Phys. Acoust.} {\bf 28},
452 (1982); T. Kambe and U. Mya-Oo,{\it J. Phys. Soc. Japan} {\bf 50},
3507 (1981).
\bibitem{fetter} An exception is the calculation of A. L. Fetter,
{\it Phys. Rev.} {\bf 136}, A1488 (1964) for the scattering of sound by a
vortex in the long wavelength approximation. In that work, long wavelength
means both $\lambda \gg a$ and $\lambda \gg \Gamma /c$. The present
paper considers the case (see text) $ \lambda \ll a$, $\lambda \sim
 \Gamma /c$.
\bibitem{deep} C. Coste and F. Lund, ``Scattering of dislocated wavefronts
by vertical vorticity and the Aharonov-Bohm effect II: Dispersive waves'', 
following paper.
\bibitem{book} L. D. Landau and E. M. Lifshitz, {\it Fluid Mechanics},
 2nd Ed., Pergamon (1987).
\bibitem{ced} Y. Aharonov and D. Bohm, {\it Phys. Rev.} {\bf 115}, 485
 (1959). For recent discussions, see R. M. Herman, {\it Found. of Phys.
} {\bf 22}, 713 (1992) and L. O'Raifeartaigh, N. Strautman and A. Wipf,
{\it Comments Nucl. Part. Phys.} {\bf 20}, 15 (1991).
\bibitem{colonius} See T. Colonius, S. K. Lele and P. Moin, {\it J. Fluid
Mech.} {\bf 260}, 271 (1994) for a treatement of this short wavelength
limit using ray-tracing methods.
\bibitem{foot1} In this case, the restriction $u \ll U$ imposed in Section
II will break down when $r$ is very small or very large. At those
points, however, the condition $u \ll c$, implicit in the derivation of
Eqn. (\ref{s19}), assures that nonlinear terms can still be neglected.
\bibitem{singularity} As noted in \cite{berry}, there is no singularity in
the forward direction. For a more recent discussion within a Born
aproximation framework, see P. V. Sakov, {\it Acoust. Phys.} {\bf 39}, 280
(1993) and R. Berthet and F. Lund,  {\it Phys. Fluids} {\bf 7},
2522 (1995).
\bibitem{vivancomelo}  F. Vivanco and F. Melo, {\it Preprint}   (1998).
\bibitem{berthet}  R. Berthet, {\it unpublished}   (1998).
\bibitem{mathematica} S. Wolfram, {\it The Mathematica Book}, Third Edition,
Cambridge University Press (1996).
\bibitem{gr} I. S. Gradshteyn and I. M. Ryshik, {\it Table of
Integrals, Series,  and Products}, Academic, 1980.
\end{references}
\end{document}